\documentstyle[preprint,aps]{revtex}
\newcommand{\bce}{\begin{center}}
\newcommand{\ece}{\end{center}}
\newcommand{\beq}{\begin{equation}}
\newcommand{\eeq}{\end{equation}}
\newcommand{\bea}{\vspace{0.25cm}\begin{eqnarray}}
\newcommand{\eea}{\end{eqnarray}}

\newcommand{\ba}{\begin{array}}
\newcommand{\ea}{\end{array}}

\newcommand{\doublespace}{
    \renewcommand{\baselinestretch}{1.6}\large\normalsize}

\def\lsim{\mathrel{\rlap{\lower4pt\hbox{\hskip1pt$\sim$}}
    \raise1pt\hbox{$<$}}}	  
\def\gsim{\mathrel{\rlap{\lower4pt\hbox{\hskip1pt$\sim$}}
    \raise1pt\hbox{$>$}}}	  

\def\beq{\begin{equation}}
\def\endeq{\end{equation}}
\def\bea{\begin{eqnarray}}
\def\arr{\begin{eqnarray}}
\def\endarr{\end{eqnarray}}

\makeindex
\begin{document}
\draft
\title{
{\bf \huge Glauber theory of initial- and final-state interactions in
$(p,2p)$ scattering\\}
}
\author{O.Benhar$^{1}$,
 S.Fantoni$^{2,3}$,
N.N.Nikolaev$^{4,5}$,
J.Speth$^{4}$,
A.A.Usmani$^{2}$,
B.G.Zakharov$^{5}$
}
\address{
$^{1}$INFN, Sezione Sanit\`{a}, Physics Laboratory,
 Istituto Superiore di Sanit\`{a}. I-00161 Roma, Italy \\
$^{2}$Interdisciplinary Laboratory, SISSA, INFN,
Sezione di Trieste. I-34014, Trieste, Italy \\
$^{3}$International Centre for Theoretical Physics,
Strada Costiera 11, I-34014, Trieste, Italy\\
$^{4}$IKP(Theorie), Forschungszentrum  J\"ulich GmbH.\\
D-52425 J\"ulich, Germany \\
$^{5}$L.D.Landau Institute for Theoretical Physics, \\
GSP-1, 117940, ul.Kosygina 2., V-334 Moscow, Russia }

\date{\today}
\maketitle
\begin{abstract}
We develop the Glauber theory description of initial- and final-state
interactions (IFSI) in quasielastic $A(p,2p)$ scattering.
We study the IFSI-distortion effects both for the inclusive
and exclusive conditions.
In inclusive reaction the important new effect is an
interaction between the
two sets of the trajectories which enter the calculation of
IFSI-distorted one-body density matrix for inclusive
$(p,2p)$ scattering and are connected
with incoherent elastic rescatterings of the initial and final protons
on spectator nucleons.
We demonstrate that IFSI-distortions of the missing
momentum distribution are large over the whole range
of missing momentum both for inclusive and exclusive reactions
and affect in a crucial way the interpretation of the
BNL data on $(p,2p)$ scattering.
Our numerical results show that in the region of
missing momentum $p_{m}\lsim 100-150$ MeV/c the incoherent IFSI
increase nuclear transparency by 5-10\%. The incoherent IFSI
become dominant at $p_{m}\gsim 200$ MeV/c.

\end{abstract}


\doublespace

\section{Introduction}
The strength of the initial- and final-state interactions (IFSI)
in $(p,2p)$ scattering is usually characterized by
the nuclear transparency, $T_{A}$, defined as a ratio
of the experimentally measured cross section to the
theoretical cross section calculated neglecting IFSI
in the plane
wave impulse approximation (PWIA).
It is expected
that, due to the color transparency (CT)
phenomenon \cite{MuellerCT,BrodskyCT}, IFSI effects will vanish
and the nuclear transparency
will tend to unity in $(p,2p)$ reaction
in the limit of
$s\rightarrow \infty$ and $|t|/s\sim 1$.
>From the point of view of the Glauber-Gribov coupled-channel multiple
scattering theory \cite{Glauber,Gribov} the vanishing of IFSI corresponds to
a cancellation of the rescattering amplitudes with
elastic (diagonal) and excited (off-diagonal) intermediate
states of the initial and final protons participating
in hard $pp$ scattering.
Naive theoretical considerations
\cite{MuellerCT,BrodskyCT}
suggest a  monotonic rise of $T_{A}$ with $s$ in the case of dominance
of the point-like perturbative mechanism of hard $pp$ scattering
\cite{BrodskyPQCD}.
However, in the BNL experiment \cite{BNL1} on large-angle $(p,2p)$
scattering at the beam momenta 6-12 GeV/c near
$\theta_{c.m.}=90^{o}$ (here $\theta_{c.m.}$ is the scattering
angle in the $pp$ center of mass frame) a decrease of $T_{A}$
was observed at beam momenta $\gsim 10$ GeV/c.
There were suggestions
\cite{BrodskyBNL,RalstonBNL,BGZBNL}
that the irregular behavior of
$T_{A}$ is due to an interplay of CT effects for hard point-like
and non-point-like, resonance or Landshoff, mechanisms of
large-angle $pp$ scattering. None the less,
from our point of view, a satisfactory explanation was not
found, and up to now the theoretical situation
is far from being clear.

In previous works on CT effects in $(p,2p)$ reaction the
IFSI-absorption effects were
treated within the optical potential approach.
This approximation
corresponds to taking into account only the coherent IFSI.
In this case the calculated cross section is related to
the exclusive $(p,2p)$ reaction, when
the final states of the residual nucleus are
exhausted by the one-hole excitations of the target nucleus.
The allowance for both the coherent and incoherent IFSI
corresponds to the inclusive reaction, when all the final
states of the residual nucleus are involved.
The recent Glauber analysis \cite{NSZ} indicates
that in the case of $(e,e'p)$ scattering the incoherent
rescatterings become dominant at high missing momenta
($\gsim 250$ MeV/c). Evidently,
in $(p,2p)$ scattering, due to the increase of the number of the fast protons
propagating through the nuclear medium as compared with
$(e,e'p)$ reaction,  the relative effect of the incoherent
rescatterings will be enhanced.
The theoretical study of the inclusive reaction would be
of great importance because the data of the BNL experiment \cite{BNL1}
correspond namely to the inclusive conditions.
The analysis of the missing momentum dependence of
the nuclear transparency in $(p,2p)$ reaction within the
coupled-channel formalism including the incoherent rescatterings
invites complications. First, evaluation of the contribution
of the off-diagonal incoherent rescatterings requires
information on the off-diagonal resonance-nucleon amplitudes
at arbitrary momentum transfer \cite{NNZ,EEPCT}. Second,
inclusion of the incoherent rescatterings make the coupled-channel
analysis complicated from the point of view of the numerical
computations.
In this situation it is reasonable
to start the study of IFSI effects in hard $(p,2p)$ reaction with
inclusion of the incoherent rescatterings within the one-channel
Glauber model.
Evidently, only after a comparison of the experimental data
with the predictions of the Glauber model one can understand
whether and to which extent the off-diagonal rescatterings
or other effects are really important.
The Glauber analysis \cite{NSZ} of the missing momentum distribution
in $(e,e'p)$ scattering shows that there is a region of
the relatively small missing
momenta ($p_{m}\lsim 150$ MeV/c) where the incoherent
rescatterings can be neglected. This fact allows one
to greatly simplify evaluation of CT effects in this region of the
missing momentum \cite{EEPCT}.
{}From the point of view of further investigations of
CT effects in $(p,2p)$ scattering it is of great importance to clarify
whether this is the case in this reaction as well.
For the above reasons the Glauber analysis of $(p,2p)$ scattering
is highly desirable. In the current literature only in ref. \cite{KYS}
the Glauber formalism was applied for evaluation of the nuclear
transparency in inclusive $(p,2p)$ reaction.
However, unjustified approximations made in ref. \cite{KYS}
led to a loss of the IFSI-distortion effects (for the criticism of the
approach
\cite{KYS} see ref. \cite{NSZ} and the discussion in section 2 of
the present paper).

In the present paper we evaluate the nuclear transparency
in $(p,2p)$ scattering within the Glauber model
in the region of
$p_{m}\lsim 300$ MeV/c.
In our analysis
we neglect the short range correlations (SRC) in the target
nucleus and describe the nucleus wave function within
independent particle shell model. In the case of the
single particle momentum distribution (SPMD) the effects
of SRC \cite{SRC} are still marginal in this region of momenta.
The analysis of $(e,e'p)$ \cite{D2,He42} shows that at $p_{m}\lsim 300$
MeV/c
SRC practically do not affect the missing momentum dependence
of the nuclear transparency as well. In $(p,2p)$ reaction the
distortion effects are enhanced as compared to the case
of $(e,e'p)$ scattering. It improves the credibility of the
independent particle shell model for analysis of
$(p,2p)$ scattering as compared to the case of $(e,e'p)$ reaction.

In our analysis, as in previous works on hard $(p,2p)$
reaction, we assume the factorization of
hard $pp$ scattering and soft IFSI. We are fully
aware that due to the strong energy dependence of the cross section
of hard $pp$ scattering ($\propto s^{-10}$) this approximation
may be questionable. The qualitative
estimates show that the off-shell effects and the nonzero energy
momentum transfer in soft IFSI in the kinematical region
of our interest
can increase
the nuclear transparency by 20-50\%.
Unfortunately a rigorous evaluation of these effects,
requiring the relativistic many body approach to
the $(p,2p)$ reaction, is not possible at present.
However, as we will see the IFSI-distortion effects are typically
much stronger than the expected magnitude of effects connected
with the off-shellness of the bound proton and
the energy transfer in soft IFSI.
Therefore, we believe that
the factorized approximation
for hard $pp$ scattering and soft IFSI
is a good starting point for evaluation of the IFSI-distortion
effects in hard $(p,2p)$ reaction.

The paper is organized as follows.
In section 2 we set out the Glauber formalism
for hard $(p,2p)$ reaction.
The numerical results are presented in section 3.
In this section we also compare the predictions of the
Glauber model with the data on the nuclear transparency
and the missing momentum distribution obtained in the BNL experiment
\cite{BNL1,BNL2}.
The summary and conclusions are presented in section 4.

\section{IFSI in \lowercase{$(p,2p)$} scattering within the Glauber formalism}
We begin with the kinematics of hard $(p,2p)$ reaction.
We denote the four momenta of the initial and two final protons
participating in hard $pp$ scattering as $(E_{1},\vec{p}_{1})$
and $(E_{3},\vec{p}_{3})$, $(E_{4},\vec{p}_{4})$  respectively.
The trajectories of the
initial and final protons will be also labeled by indexes 1 and 3,4.
We use the coordinate system with $z$-axis along $\vec{p}_{1}$ and
the $x$-axis in the reaction plane.
At high energy for $\theta_{c.m}=90^{o}$ the angle between
the three momenta of the final protons and $z$-axis in the
laboratory frame, $\theta_{l.f.}$, becomes small
($\theta_{l.f.}\sim [2m_{p}(E_{1}-m_{p})]^{1/2}/E_{1}$, here
$m_{p}$ is the proton mass). Below we will make use of this fact to
simplify the numerical calculations.
As was stated in
section 1, we assume the factorization of hard $pp$ scattering
and soft IFSI of the fast protons with spectator nucleons.
Then, the differential cross section of $(p,2p)$ scattering
can be written through the distorted spectral function,
$S(E_{m},p_{m},E_{1})$,
in the form
\beq
\frac{d^{\,3}\sigma}{dtdE_{m}d\vec{p}^{\,\,3}_{m}}
(E_{1},t,E_{m},\vec{p}_{m})=
\frac{d\sigma_{pp}}{dt}(s,t)S(E_{m},\vec{p}_{m},E_{1})\,,
\label{eq:1.1}
\eeq
where $d\sigma_{pp}/dt$ is the differential
cross section of hard $pp$ scattering,
the missing momentum and missing energy are defined as
$\vec{p}_{m}=\vec{p}_{3}+\vec{p}_{4}-\vec{p}_{1}$,
$E_{m}=E_{3}+E_{4}-E_{1}$,
$s$ is the center of mass energy squared of the $pp$ system.
To leading order in the missing momentum $s$ is given by
\beq
s\approx s_{0}\left(1-\frac{p_{m,z}}{m_{p}}\right)\,,
\label{eq:1.1p}
\eeq
where $s_{0}=2m_{p}(E_{1}+m_{p})$. Notice that
keeping in Eq.~(\ref{eq:1.1p})
the second order terms in $\vec{p}_{m}$ does not
make much sense because the energy momentum
transfer in the soft rescatterings of the fast initial and
final protons in the nuclear medium, which is neglected in
the factorized approximation (\ref{eq:1.1}), also gives the effect
of the second order in $\vec{p}_{m}$. As was mentioned in section 1,
an accurate treatment of such
effects, and of the off-shell effects,
requires making use of the relativistic many body
approach, which goes beyond the scope of our exploratory study.

The distorted spectral function, which, under the factorized
approximation (\ref{eq:1.1}), accumulates all the IFSI effects,
can be written as
\beq
S(E_{m},\vec{p}_{m})=\sum_{f}|M_{f}(\vec{p}_{m})|^{2}
\delta(E_{m}+E_{A-1}(\vec{p}_{m})-m_{A})\,,
\label{eq:1.2}
\eeq
where
$M_{f}(\vec{p}_{m})$ is the reduced matrix element of the
exclusive process
 $p\,+\,A_{i}\,\rightarrow\,\,p\,+\,p+\,(A-1)_{f}$.
In Eq.~(\ref{eq:1.2}) and hereafter for the sake
of brevity the
variable $E_{1}$  is suppressed.
In the present paper we will study IFSI-distortion effects
at the level of the missing momentum distribution, $w(\vec{p}_{m})$,
which reads
\beq
w(\vec{p}_{m})={1 \over (2\pi)^{3}} \int
dE_{m} S(E_{m},\vec{p}_{m})\,.
\label{eq:1.3}
\eeq
The substitution of (\ref{eq:1.2}) into (\ref{eq:1.3}) yields
\beq
w(\vec{p}_{m})=
{1\over (2\pi)^{3}}
\sum_{f}|M_{f}(\vec{p}_{m})|^{2}\,.
\label{eq:1.4}
\eeq

In our analysis we confine ourselves to a large mass
number of the target nucleus $A\gg 1$. Then, neglecting
the center of mass correlations we can write
$M_{f}(\vec{p}_{m})$ as
\beq
M_{f}(\vec{p}_{m})=\int d^{3}\vec{r}_{1}...d^{3}\vec{r}_{A}
\Psi_{f}^{*}(\vec{r}_{2},...,\vec{r}_{A})
\Psi_{i}(\vec{r}_{1},...,\vec{r}_{A})
S(\vec{r}_{1},...\vec{r}_{A})\exp(i\vec{p}_{m}\vec{r}_{1})\,.
\label{eq:1.5}
\eeq
Here $\Psi_{i}$ and $\Psi_{f}$ are wave functions
of the target and residual nucleus, respectively. The nucleon "1" is
chosen to be the struck proton. The use of the same indexes
for the spatial coordinates of the nucleons in the nucleus wave functions
and for labeling the fast protons participating in hard $pp$ scattering
must not lead to a confusion because we will not use the spatial
coordinates of the fast initial and final protons.
For the sake of brevity, in Eq.~(\ref{eq:1.5}) and hereafter
the spin and isospin variables are suppressed.
The factor $S(\vec{r}_{1},...,\vec{r}_{A})$ in Eq.~(\ref{eq:1.5})
takes into account the soft IFSI-distortion effects. In the Glauber
model it can be written in the form
\beq
S(\vec{r}_{1},...,\vec{r}_{A})=
S_{1}(\vec{r}_{1},...,\vec{r}_{A})
S_{3}(\vec{r}_{1},...,\vec{r}_{A})
S_{4}(\vec{r}_{1},...,\vec{r}_{A})
\label{eq:1.6}
\eeq
where $S_{1,3,4}(\vec{r}_{1},...,\vec{r}_{A})$ are the
absorptive
factors for the initial and
final protons which are given by
\beq
S_{i}(\vec{r}_{1},...,\vec{r}_{A})=\prod\limits_{j=2}^{A}
\left[1-\gamma_{i}(\vec{r}_{1},\vec{r}_{j})\right]\,,
\label{eq:1.7}
\eeq
with
\beq
\gamma_{i}(\vec{r}_{1},\vec{r}_{j})=
\theta(\vec{n}_{i}(\vec{r}_{j}-\vec{r}_{1}))
\Gamma_{i}(b_{i}(\vec{r}_{1},\vec{r}_{j}))\,.
\label{eq:1.8}
\eeq
Here $\vec{n}_{i}$ are the unit
vectors defined as $\vec{n}_{1}=-\vec{p}_{1}/|\vec{p}_{1}|$
and $\vec{n}_{3,4}=\vec{p}_{3,4}/|\vec{p}_{3,4}|$,
$b_{i}(\vec{r}_{1},\vec{r}_{j})=\left[(\vec{r}_{j}-\vec{r}_{1})^{2}-
(\vec{n}_{i}(\vec{r}_{j}-\vec{r}_{1}))^{2}\right]^{1/2}\,\,$ is
the transverse distance between the spectator nucleons
"j" and the trajectory of the fast (initial or final) proton "i",
$\Gamma_{i}$ is the familiar profile function of the elastic
proton-nucleon scattering (the label "i" reflects the fact that
the profile function must be calculated at the energy $E_{i}$).
Eqs. (\ref{eq:1.6}), (\ref{eq:1.7}) are written under the usual
assumption that the spectator
coordinates can be considered as frozen during propagation of
the fast protons through the nuclear medium. Also, we neglect
the interaction radius of $90^{o}$ hard $pp$ scattering, which is
expected to be $\sim 1/\sqrt{s}$.

In our calculations we use for $\Gamma(\vec{b})$ the standard
high-energy parameterization
\beq
\Gamma(b)={\sigma_{tot}(pN) (1-i\alpha_{pN})\over 4\pi B_{pN}}
\exp\left[-{b^{2}\over 2B_{pN}}\right] \,.
\label{eq:1.9}
\eeq
Here $\alpha_{pN}$ is the ratio of the real to imaginary part
of the forward elastic $pN$ amplitude, $B_{pN}$ is the diffractive
slope describing the $t$ dependence of the elastic proton-nucleon
cross section
\beq
\frac{d\sigma_{el}(pN)}{dt}=
\left.\frac{d\sigma_{el}(pN)}{dt}\right|_{t=0}
\exp(-B_{pN}|t|)\,.
\label{eq:1.10}
\eeq

After substituting expression (\ref{eq:1.5}) into Eq.~(\ref{eq:1.4})
and making use
of the closure relation
\beq
\sum_{f}\Psi_{f}(\vec{r}_{2}^{\,'},...,\vec{r}_{A}^{\,'})
\Psi^{*}_{f}(\vec{r}_{2},...,\vec{r}_{A})=
\prod\limits_{j=2}^{A}\delta(\vec{r}_{j}-\vec{r}_{j}^{\,'})
\label{eq:1.11}
\eeq
to sum
over all the final states of the residual nucleus,
we can represent $w(\vec{p}_{m})$ in the form
\beq
w(\vec{p}_{m})=\frac{1}{(2\pi)^{3}}\int
d^{3}\vec{r}_{1}d^{3}\vec{r}_{1}^{\,'}
\rho_{D}(\vec{r}_{1},\vec{r}_{1}^{\,'})
\exp[i\vec{p}_{m}(\vec{r}_{1}-\vec{r}_{1}^{\,'})]\,,
\label{eq:1.12}
\eeq
where
\beq
\rho_{D}(\vec{r}_{1},\vec{r}_{1}^{\,'})=
\int
\prod\limits_{j=2}^{A}d^{3}\vec{r}_{j}
\Psi_{i}(\vec{r}_{1},\vec{r}_{2},...,\vec{r}_{A})
\Psi_{i}^{*}(\vec{r}_{1}^{\,'},\vec{r}_{2},...,\vec{r}_{A})
S(\vec{r}_{1},\vec{r}_{2},...,\vec{r}_{A})
S^{*}(\vec{r}_{1}^{\,'},\vec{r}_{2},...,\vec{r}_{A})\,.
\label{eq:1.13}
\eeq
 The function $\rho_{D}(\vec{r}_{1},\vec{r}_{1}^{\,'})$
may be viewed as an IFSI-modified one-body proton density matrix.
In the PWIA, when  the IFSI factors in the right-hand side
of Eq.~(\ref{eq:1.13}) equal unity, it reduces to
the formula  for the usual one-body
proton density matrix $\rho(\vec{r}_{1},\vec{r}_{1}^{\,'})$,
and Eq.~(\ref{eq:1.12}) reduces to the expression for SPMD
\beq
n_{F}(\vec{p}_{m})
=
{1\over (2\pi)^{3}}\int
 d\vec{r}_{1}d\vec{r}_{1}^{\,'}
 \rho(\vec{r}_{1},\vec{r}_{1}^{\,'})
\exp\left[i\vec{p}_{m}(\vec{r}_{1}-\vec{r}_{1}^{\,'})\right] \,.
\label{eq:1.14}
\eeq

As was stated in section 1, we will
describe the target nucleus in the
independent particle shell model. After neglecting the SRC
the $A$-body semidiagonal density matrix
$\Psi_{i}(\vec{r}_{1},\vec{r}_{2},...,\vec{r}_{A})
\Psi_{i}^{*}(\vec{r}_{1}^{\,'},\vec{r}_{2},...,\vec{r}_{A})$
still contains the Fermi correlations.
To carry out the integration over the coordinates of the
spectator nucleons in Eq.~(\ref{eq:1.13}) we neglect the Fermi
correlations
and replace the $A$-body semidiagonal density matrix by the
factorized form
\beq
\Psi_{i}(\vec{r}_{1},\vec{r}_{2},...,\vec{r}_{A})
\Psi_{i}^{*}(\vec{r}_{1}^{\,'},\vec{r}_{2},...,\vec{r}_{A})
\rightarrow
\rho(\vec{r}_{1},\vec{r}_{1}^{\,'})\prod\limits_{i=2}^{A}
\rho(\vec{r}_{i})
\,.
\label{eq:1.15}
\eeq
Here
$$
\rho(\vec{r}_{1}, \vec{r}_{1}^{\,'})=\frac{1}{Z} \sum_{n}
\phi_{n}^{*}(\vec{r}_{1}^{\,'})\phi_{n} ( \vec{r}_{1})
$$
is the shell model one-body proton density matrix and
$\phi_{n}$ are the shell model wave functions,
$\rho_{A}(\vec{r}\,)$ is the normalized to unity nucleon nuclear density.
The errors connected with ignoring the Fermi correlations
must be small because
the ratio between the Fermi correlation length $l_{F}\sim 3/k_{F}$
and the interaction length corresponding to the interaction of the
fast initial and final protons with the Fermi correlated
spectator nucleons
$l_{int}\sim 4(\sigma_{tot}(pN)\langle n_{A}\rangle)^{-1}$
(here $\langle n_{A}\rangle$ is the average nucleon nuclear density)
is a small quantity ($\sim 0.25$).
Recall, that a high accuracy of the factored approximation for
the many-body nuclear density in the
calculation of the Glauber model attenuation factor for
the small angle hadron-nucleus scattering is
well known for a long time (for an extensive review on $hA$
scattering see \cite{Alkhaz}).

After making use of the replacement (\ref{eq:1.15}) in
Eq.~(\ref{eq:1.13}) the missing momentum
distribution (\ref{eq:1.12}) can be written as follows
\beq
w(\vec{p}_{m})=\frac{1}{(2\pi)^{3}}\int
d^{3}\vec{r}_{1}d^{3}\vec{r}_{1}^{\,'}
\rho(\vec{r}_{1},\vec{r}_{1}^{\,'})
\Phi(\vec{r}_{1},\vec{r}_{1}^{\,'})
\exp[i\vec{p}_{m}(\vec{r}_{1}-\vec{r}_{1}^{\,'})]\,,
\label{eq:1.16}
\eeq
where
the IFSI factor $\Phi(\vec{r}_{1},\vec{r}_{1}^{\,'})$
is given by
\bea
\Phi(\vec{r}_{1},\vec{r}_{1}^{\,'})=
\int \prod\limits_{j=2}^{A}\rho_{A}(\vec{r}_{j}) d^{3}\vec{r}_{j}
S(\vec{r}_{1},\vec{r}_{2},...,\vec{r}_{A})
S^{*}(\vec{r}_{1}^{\,'},\vec{r}_{2},...,\vec{r}_{A})\nonumber\\
=\left[\int d^{3}\vec{r}
\rho_{A}(\vec{r}\,)P(\vec{r}_{1},\vec{r}_{1}^{\,'},\vec{r}\,)
\right]^{A-1}\;,\;\;\;\;\;\;\;\;\;\;
\label{eq:1.17}
\eea
with
\bea
P(\vec{r}_{1},\vec{r}_{1}^{\,'},\vec{r}\,)=
[1-\gamma_{1}(\vec{r}_{1},\vec{r}\,)]
[1-\gamma_{3}(\vec{r}_{1},\vec{r}\,)]
[1-\gamma_{4}(\vec{r}_{1},\vec{r}\,)]\nonumber\\
\times
[1-\gamma_{1}^{*}(\vec{r}_{1}^{\,'},\vec{r}\,)]
[1-\gamma_{3}^{*}(\vec{r}_{1}^{\,'},\vec{r}\,)]
[1-\gamma_{4}^{*}(\vec{r}_{1}^{\,'},\vec{r}\,)]\,\,.
\label{eq:1.18}
\eea

The right-hand side of Eq.~(\ref{eq:1.18}) contains the terms
up to sixth order in the profile functions. To simplify the
calculations we neglect the terms which
contain the products
$\gamma_{1}(\vec{r}_{1},\vec{r}\,)\gamma_{3,4}(\vec{r}_{1},\vec{r}\,)$
and
$\gamma_{1}^{*}(\vec{r}_{1}^{\,'},\vec{r}\,)
\gamma_{3,4}^{*}(\vec{r}_{1}^{\,'},\vec{r}\,)$.
Such terms
correspond to simultaneous interactions of the spectator
nucleon with the initial and final protons. Due to the above
mentioned smallness
of the scattering angle for hard $pp$ reaction in the laboratory
frame
(at $p_{1}\sim 10$ GeV/c  $\,\,\,\theta_{l.f.}\sim 25^{o}$) they
are only important in a narrow vicinity of the spectator
position $\vec{r}$ with the longitudinal extension considerably smaller
than the interaction length of fast protons in the nuclear
medium.
For
this reason these terms can be safely neglected in calculating the
IFSI factor (\ref{eq:1.17}). It is also worth noting that a
rigorous treatment
of such effects requires
taking into account the quark content of the proton and can not
be performed within the Glauber model.

After neglecting
the simultaneous interactions of the spectators with the initial and
final protons
the IFSI factor
$\Phi(\vec{r}_{1},\vec{r}_{1}^{\,'})$
can be written in the form
\bea
\Phi(\vec{r}_{1},\vec{r}_{1}^{\,'})
=\left\{1+\frac{1}{A}\left[\sum\limits_{i=1,3,4}G_{i}(\vec{r}_{1})
+\sum\limits_{i=1,3,4}G^{i}(\vec{r}_{1}^{\,'})
+G_{34}(\vec{r}_{1})+G^{34}(\vec{r}_{1}^{\,'})\right.\right.
\nonumber\\ \left.\left.
+\sum\limits_{i,j=1,3,4}G_{i}^{j}(\vec{r}_{1},\vec{r}_{1}^{\,'})
+\sum\limits_{i=1,3,4}G_{34}^{i}(\vec{r}_{1},\vec{r}_{1}^{\,'})
+\sum\limits_{i=1,3,4}G_{i}^{34}(\vec{r}_{1},\vec{r}_{1}^{\,'})
+G_{34}^{34}(\vec{r}_{1},\vec{r}_{1}^{\,'})\right]\right\}^{A-1}
\,,
\label{eq:1.19}
\eea

where
\bea
G_{i_{1},...,i_{n}}^{j_{1},...,j_{m}}(\vec{r}_{1},\vec{r}_{1}^{\,'})
=(-1)^{n+m}\int d^{3}\vec{r}n_{A}(\vec{r}\,)
\gamma_{i_{1}}(\vec{r}_{1},\vec{r}\,)...
\gamma_{i_{n}}(\vec{r}_{1},\vec{r}\,)
\gamma_{j_{1}}^{*}(\vec{r}_{1}^{\,'},\vec{r}\,)...
\gamma_{j_{m}}^{*}(\vec{r}_{1}^{\,'},\vec{r}\,)\nonumber\\=
(-1)^{n+m}\int d^{3}\vec{r}n_{A}(\vec{r}\,)
\theta(\vec{n}_{i_{1}}(\vec{r}-\vec{r}_{1}))...
\theta(\vec{n}_{i_{n}}(\vec{r}-\vec{r}_{1}))
\theta(\vec{n}_{j_{1}}(\vec{r}-\vec{r}_{1}^{\,'}))...
\theta(\vec{n}_{j_{m}}(\vec{r}-\vec{r}_{1}^{\,'}))
\nonumber\\ \times
\Gamma_{i_{1}}(b_{i_{1}}(\vec{r}_{1},\vec{r}\,))...
\Gamma_{i_{n}}(b_{i_{n}}(\vec{r}_{1},\vec{r}\,))
\Gamma_{j_{1}}^{*}(b_{i_{1}}(\vec{r}_{1}^{\,'},\vec{r}\,))...
\Gamma_{j_{m}}^{*}(b_{j_{m}}(\vec{r}_{1}^{\,'},\vec{r}\,))
\label{eq:1.20}
\eea
(here $n_{A}(\vec{r}\,)=A\rho_{A}(\vec{r}\,)$ is the nuclear density).

The IFSI factor (\ref{eq:1.19})
corresponds to the inclusive $(p,2p)$ reaction, when all the
final states of the residual nucleus are allowed.
In a similar way, starting from the matrix element (\ref{eq:1.5}) and taking
into account in the sum over the final states of the residual
nucleus in Eq.~(\ref{eq:1.4}) only the one-hole hole excitations of
the target nucleus, one can obtain the coherent IFSI factor for
the exclusive reaction
\beq
\Phi_{coh}(\vec{r}_{1},\vec{r}_{1}^{\,'})=
S_{coh}(\vec{r}_{1})S_{coh}^{*}(\vec{r}_{1}^{\,'})\,,
\label{eq:1.22}
\eeq
where
\beq
S_{coh}(\vec{r}_{1})=
\left\{1+\frac{1}{A}\left[
\sum\limits_{i=1,3,4}G_{i}(\vec{r}_{1})
+G_{34}(\vec{r}_{1})\right]\right\}^{A-1}\,.
\label{eq:1.22p}
\eeq
The factorized form of $\Phi_{coh}(\vec{r}_{1},\vec{r}_{1}^{\,'})$
allows one to write the
missing momentum distribution for exclusive reaction,
which we will refer to as $w_{exc}(\vec{p}_{m})$,
as a sum of the IFSI-distorted distributions for the one-hole
excitations
\beq
w_{exc}(\vec{p}_{m})=
\frac{1}{Z}\sum\limits_{n}
\left|\int d^{3}\vec{r}_{1}\phi_{n}(\vec{r}_{1})\exp(i\vec{p}_{m}\vec{r}_{1})
S_{coh}(\vec{r}_{1})
\right|^{2}\,.
\label{eq:1.23}
\eeq

The terms $G_{i}$
in (\ref{eq:1.22p}) describe the usual
attenuation of the initial and final protons in
the nuclear medium, while the term $G_{34}$
is related to the shadowing effect in the system of the final protons.
It is a connected with the rescatterings of the protons "3" and "4"
on the same spectator nucleon. The transverse separation of
the trajectories "3" and "4" is $\sim 2\theta_{l.f.}(z-z_{1})$
(here $z$ is the longitudinal coordinate of the spectator nucleon).
Hence, the simultaneous interaction of the spectator nucleon
with both fast final protons is possible
up to the longitudinal distance $\sim R_{int}/\theta_{l.f.}$
(here $R_{int}=\sqrt{2B(pN)}\approx 0.8$  $fm$ is the interaction
radius for soft $pN$-scattering) from the hard collision vertex.
At incident beam momentum
$\sim 10$ GeV/c this size becomes
as large as the
absorption length for the final protons in the nuclear
medium. None the less, as we will see in the energy region  of
the BNL experiment \cite{BNL1}
the shadowing
correction
to the nuclear transparency turns out to be relatively small.

Let us turn to the whole IFSI factor (\ref{eq:1.19}).
The difference between $\Phi(\vec{r}_{1},\vec{r}_{1}^{\,'})$
and $\Phi_{coh}(\vec{r}_{1},\vec{r}_{1}^{\,'})$ is connected
with the soft incoherent rescatterings of the fast initial
and final protons in the nuclear medium.
The emergence of the interference
terms, like $G_{i}^{j}$ with
$i\neq j$, and of the shadowing terms $G_{34}^{i}$, $G_{i}^{34}$
and $G_{34}^{34}$ in Eq.~(\ref{eq:1.19}) shows that the process of
excitation of the
residual nucleus can not be considered as a simple superposition of
the effects produced
by the initial and final protons. It makes impossible a
probabilistic interpretation of the incoherent IFSI.
The incoherent rescatterings  make
the IFSI factor (\ref{eq:1.19}) a nonfactorized function of
$\vec{r}_{1}$ and $\vec{r}_{1}^{\,'}$.
To leading order in $1/A$ the nonfactorized form of the IFSI
factor for inclusive reaction is connected with
an interaction between the two sets of the trajectories
(originating from $\vec{r}_{1}$ and $\vec{r}_{1}^{\,'}$)
generated by the overlapping of the functions
$\gamma_{i}(\vec{r}_{1},\vec{r}\,)$ and
$\gamma_{j}^{*}(\vec{r}_{1}^{\,'},\vec{r}\,)$ in the integral
(\ref{eq:1.20}) for $G$-functions with $i_{n},j_{m}\ge 1$.
Due to the fast decrease of the profile function
at $b\gsim R_{int}$,
the integral (\ref{eq:1.20}), and the interaction between the two
sets of the trajectories, vanishes unless there is a region
where $\vec{r}$ is close to all
the trajectories simultaneously which emerge in (\ref{eq:1.20})
through the functions
$\gamma_{i}(\vec{r}_{1},\vec{r}\,)$ and
$\gamma_{j}^{*}(\vec{r}_{1}^{\,'},\vec{r}\,)$.
Simple geometrical consideration shows that
the interaction between the two sets of the
trajectories generated by the incoherent IFSI
is for the most part important in the region of
$|y_{1}-y_{1}^{\,'}|\lsim R_{int}$.
In the variables $(x_{1}-x_{1}^{'})$ and $(z_{1}-z_{1}^{'})$
this interaction vanishes more slowly and survive  at
distances
$\gsim R_{int}$ as well. Evidently, as in the case of $(e,e'p)$ reaction
\cite{NSZ}, the short range interaction
between the two sets of the trajectories at the level of
the Glauber absorptive factor will, for the most part,
affect the missing momentum distribution at
$p_{m}\gsim 1/R_{int}\sim 200$ MeV/c. However, on the
contrary to $(e,e'p)$ reaction, in $(p,2p)$ scattering
due to the geometry of hard $pp$ scattering, the
azimuthal symmetry in the $(x,y)$ plane is absent.
The effect of the incoherent rescatterings
must be enhanced in the region of large $p_{m,y}$
as compared with the cases of large $p_{m,x}$
or $p_{m,z}$.

The numerical calculations can be simplified exploiting the fact that
in the integral (\ref{eq:1.20}) $n_{A}(\vec{r}\,)$ is a smooth
function (in transverse directions) as compared to the profile
functions.
To factor out in (\ref{eq:1.20}) the transverse integration,
we introduce a new coordinate
system with the $z$-axis defined as a line which is
the center of "gravity" of the trajectories
$i_{1},...,i_{n},j_{1},...,j_{m}$ in the $(x,y)$ plane. Then,
in terms of the new longitudinal,  $\xi$, and transverse,
$\vec{\tau}$, variables, to leading
order in the small parameter $R_{int}^{2}/R_{A}^{2}$ ( $R_{A}$ is the
nucleus radius) the integral (\ref{eq:1.20}) can be written as
\bea
G_{i_{1},...,i_{n}}^{j_{1},...,j_{m}}(\vec{r}_{1},\vec{r}_{1}^{\,'})
=\int d \xi\,
n_{A}(0,0,\xi)
\theta(\vec{n}_{\xi}(\vec{r}-\vec{r}_{1}))...
\theta(\vec{n}_{\xi}(\vec{r}-\vec{r}_{1}))
\theta(\vec{n}_{\xi}(\vec{r}-\vec{r}_{1}^{\,'}))
\theta(\vec{n}_{\xi}(\vec{r}-\vec{r}_{1}^{\,'}))\nonumber\\
\times
\int d^2\vec{\tau}\;
\Gamma_{i_{1}}(b_{i_{1}}(\vec{r}_{1},\vec{r}\,))...
\Gamma_{i_{n}}(b_{i_{n}}(\vec{r}_{1},\vec{r}\,))
\Gamma_{j_{1}}^{*}(b_{j_{1}}(\vec{r}_{1}^{\,'},\vec{r}\,))...
\Gamma_{j_{m}}^{*}(b_{j_{m}}(\vec{r}_{1}^{\,'},\vec{r}\,))\,.
\;\;\;\;\;\;\;\;
\label{eq:1.25}
\eea
Here we have written the nuclear density $n_{A}$ as a function
of the new variables, and the vector $\vec{r}$ must be
treated as a vector-function of $\xi$ and $\vec{\tau}$.
In Eq.~(\ref{eq:1.25}) we have also taken advantage of the smallness of
$\theta_{l.f.}$ and made replacements
$\vec{n}_{i,j}\rightarrow \vec{n}_{\xi}$, where $\vec{n}_{\xi}$
is the unit vector along  $\xi$-axis. Evidently,
similar to neglecting the
simultaneous interaction of the spectator nucleon with
the initial and final fast protons,
such a replacement
spoils the Glauber form of the attenuation factor only in a very
narrow region of the spectator positions
in the vicinity of the hard $pp$ interaction
vertex and practically does not affect the final numerical
results. The integral over the transverse coordinates in
Eq.~(\ref{eq:1.25})
has the Gaussian form and can be calculated
analytically. The corresponding formulas are somewhat lengthy and
we do not present them here.
The remaining integration over $\xi$ was carried out
numerically.

Eqs.~(\ref{eq:1.16}),~(\ref{eq:1.19}),
~(\ref{eq:1.25})
form the basis
for evaluation of the missing momentum distribution
$(p,2p)$ reaction within the Glauber model.
Then, the nuclear transparency
for a certain kinematical domain, $D$, of the missing momentum
can be calculated through the formula
\beq
T_{A}(D)=\frac{\int\limits_{D} d^{3}\vec{p}_{m}
w(\vec{p}_{m})}
{\int\limits_{D} d^{3}\vec{p}_{m}
n_{F}(\vec{p}_{m})}\,.
\label{eq:1.25p}
\eeq
Besides the calculations of the nuclear transparency
as a function of $\vec{p}_{m}$ for the point-like domain $D$
$T_{A}(\vec{p}_{m})=w(\vec{p}_{m})/n_{F}(\vec{p}_{m})$,
in the present paper we calculate
the nuclear transparency for the kinematical domains including
all the values of the two and three components of the missing momentum.
The one-dimensional missing momentum distributions obtained after
the integration
of $w(\vec{p}_{m})$ over the two components of $\vec{p}_{m}$
are given by
\beq
w_{x}(p_{m,x})=\frac{1}{2\pi}
\int dx dx^{'}dydz\rho(x,y,z,x^{'},y,z)
\Phi(x,y,z,x^{'},y,z)
\exp[i\vec{p}_{m,x}(x-x^{'})]\,,
\label{eq:1.26}
\eeq
\beq
w_{y}(p_{m,y})=\frac{1}{2\pi}
\int dx dy dy^{'}dz\rho(x,y,z,x,y^{'},z)
\Phi(x,y,z,x,y^{'},z)
\exp[i\vec{p}_{m,y}(y-y^{'})]\,,
\label{eq:1.27}
\eeq
\beq
w_{z}(p_{m,z})=\frac{1}{2\pi}
\int dxdydz dz^{'}\rho(x,y,z,x,y,z^{'})
\Phi(x,y,z,x,y,z^{'})
\exp[i\vec{p}_{m,z}(z-z^{'})]\,,
\label{eq:1.28}
\eeq

The integrated nuclear
transparency corresponding to the whole kinematical domain of the missing
momentum in Eq.~(\ref{eq:1.25p}) is given by
\beq
T_{A}=\int d^{3}\vec{r}_{1}\rho_{A}(\vec{r}_{1})
\Phi(\vec{r}_{1},\vec{r}_{1})\,.
\label{eq:1.29}
\eeq
Making use of (\ref{eq:1.19}), (\ref{eq:1.25})
after a simple algebra
we can represent (\ref{eq:1.29}) as
\beq
T_{A}=\int d^{3}\vec{r}_{1}\rho_{A}(\vec{r}_{1})
\left\{1-\frac{1}{A}\left[\sum\limits_{i=1,3,4}\sigma_{in}^{pN}(E_{i})
t(\vec{r}_{1},\vec{n}_{i})
-\delta(\vec{r}_{1})\right]\right\}^{A-1}\,,
\label{eq:1.30}
\eeq
where
$$
t(\vec{r}_{1},\vec{n}_{i})=\int\limits_{0}^{\infty}
dl n_{A}(\vec{r}_{1}+\vec{n}_{i}l)
$$
is the partial optical thickness function, and
\bea
\delta(\vec{r}_{1})=
G_{34}(\vec{r}_{1})+G^{34}(\vec{r}_{1})+
G_{3}^{4}(\vec{r}_{1},\vec{r}_{1})+
G_{4}^{3}(\vec{r}_{1},\vec{r}_{1})
+G_{34}^{3}(\vec{r}_{1},\vec{r}_{1})
+G_{34}^{4}(\vec{r}_{1},\vec{r}_{1})\nonumber\\+
G_{34}^{3}(\vec{r}_{1},\vec{r}_{1})+
G_{34}^{4}(\vec{r}_{1},\vec{r}_{1})+
G_{34}^{34}(\vec{r}_{1},\vec{r}_{1})
\,.\;\;\;\;\;\;\;\;\;\;\;\;\;\;\;\;
\label{eq:1.32}
\eea
The terms containing the inelastic $pN$ cross sections
in the square brackets in the right-hand side of (\ref{eq:1.30}) describe
the
usual absorption in propagation of the fast protons in the nuclear
medium, while $\delta(\vec{r}_{1})$ yields correction
for the shadowing in the final $pp$-system
and the interference effects in the incoherent IFSI.
Our numerical calculations show that
this correction is non-negligible.
Thus,
even in the case of the integrated nuclear transparency, the IFSI do
not allow
a probabilistic interpretation. However, it is worth noting,
that the integrated nuclear transparency, as well as the
transverse missing momentum distributions
(\ref{eq:1.26}), (\ref{eq:1.27}),
are not affected
by the interference of amplitudes for rescatterings of the initial
and final protons. One can see from Eq.~(\ref{eq:1.25})
that the $G$-functions in the IFSI factor
(\ref{eq:1.19}) related to this interference
vanish for $z_{1}=z_{1}^{'}$.

Eq.~(\ref{eq:1.30}) yields the
nuclear transparency for the inclusive $(p,2p)$ reaction.
In the case of the exclusive reaction, after substituting
in (\ref{eq:1.29}) the coherent IFSI factor
(\ref{eq:1.22}), one can obtain for the integrated nuclear
transparency
\bea
T_{A}^{exc}=\int d^{3}\vec{r}_{1}\rho_{A}(\vec{r}_{1})
\left\{1-\frac{1}{A}
\left[\frac{1}{2}\sum\limits_{i=1,3,4}
\sigma_{tot}^{pN}(E_{i})\left(1-i\alpha_{pN}(E_{i})\right)
t(\vec{r}_{1},\vec{n}_{i})-
G_{34}(\vec{r}_{1})
\right]\right\}^{A-1}\nonumber\\
\times
\left\{1-\frac{1}{A}\left[\frac{1}{2}\sum\limits_{i=1,3,4}
\sigma_{tot}^{pN}(E_{i})\left(1+i\alpha_{pN}(E_{i})\right)
t(\vec{r}_{1},\vec{n}_{i})-
G_{34}^{*}(\vec{r}_{1})
\right]\right\}^{A-1}.\;\;\;\;\;\;\;\;\;
\label{eq:1.32p}
\eea

We conclude this section with a short comment on the
work \cite{KYS} which previously considered
$(p,2p)$ scattering within the Glauber model.
The authors of \cite{KYS} also assume the
factorization of hard $pp$ scattering and soft IFSI, and use the
factorized approximation (\ref{eq:1.15}) for
the $A$-body semidiagonal density matrix.
However, then
they make use of the approximations
which can not be justified.
First, they neglect in
their counterpart of our equation (\ref{eq:1.16})
the dependence of the IFSI factor on
$\Delta \vec{r}_{1}=(\vec{r}_{1}-\vec{r}_{1}^{\,'})$, and
put
$
\vec{r}_{1}=\vec{r}_{1}^{\,'}=(\vec{r}_{1}+\vec{r}_{1}^{\,'})/2\,.
$
Under this approximation for the factorized parametrization
of the one-body proton density matrix
$
\rho(\vec{r},\vec{r}^{\,'})=\rho_{A}
(\frac{1}{2}(\vec{r}+\vec{r}^{\,'}))W(\vec{r}-\vec{r}^{\,'})
$
(here $W(\vec{r}-\vec{r}^{\,'})$ is the Fourier transform of
the SPMD) they obtained the
missing momentum distribution which is proportional to the SPMD.
Evidently, the approach of ref. \cite{KYS} misses
all the distortion effects which, as we shall demonstrate
below, are quite strong.
Our predictions for the integrated
nuclear transparency also differ from the results of ref. \cite{KYS},
because in the analysis \cite{KYS} the shadowing in the final
$pp$-system and interference effects in the incoherent IFSI
were not taken into account.

\section{Numerical results}

In this section we present our numerical results based
on the formalism developed in the previous section.
We performed the calculations
for the target nuclei $^{12}C$, $^{16}O$, $^{27}Al$ and $^{40}Ca$.
For $^{12}C$ and $^{27}Al$ we compare the numerical results with
the available data of the BNL experiment \cite{BNL1,BNL2}.
The calculations were performed for the oscillator shell wave
functions.
We adjusted the oscillator frequency, $\omega_{osc}$, for the above
set
of the nuclei to reproduce the experimental values of the
root-mean-square radius of the charge distribution:
$\langle r^{2}\rangle^{1/2}$=
2.47,
2.73,
3.05 and
3.47 $fm$ for
$^{12}C$,
$^{16}O$,
$^{27}Al$ and
$^{40}Ca$
\cite{Atdata}, respectively.
We obtained the
following set of the values of the oscillator
radius $r_{osc}=(m_{p}\omega_{osc})^{-1/2}=1.59$,
1.74, 1.78 and 1.95 $fm$.
The difference between the charge distribution and the proton
nuclear density connected with the proton charge radius was
taken into account.
The charge density and SPMD
in the region of $p_{m}\lsim 300$ Mev/c, calculated with
the above set of the oscillator radii
are
practically
indistinguishable from
the results of more involved
Hartree-Fock calculations.
The SPMD calculated in the harmonic oscillator shell model is
also close to the one obtained within
a many-body approach with realistic nucleon-nucleon potential
in ref. \cite{BCS}.

As it was stated in section 2,
we use the exponential parameterization
of the proton-nucleon elastic amplitude. The diffraction slope of the
$pN$ scattering was estimated from the relation
\beq
B_{pN}\approx \frac{\sigma_{tot}^{2}(pN)(1+\alpha_{pN}^{2})}
{16\pi \sigma_{el}(pN)}\,.
\label{eq:2.1}
\eeq
In our calculations we define the $pN$ cross sections and
$\alpha_{pN}$ as mean values of these quantities for the
$pp$ and $pn$ scattering. We borrowed the experimental data
on $pp$, $pn$ cross sections and $\alpha_{pp}$, $\alpha_{pn}$ from
the recent review \cite{Lehar}.

In Figs.~1,~2 and 3 we show the nuclear transparency
for $^{16}O(p,2p)$ and $^{40}Ca(p,2p)$ reactions
at $p_{lab}=6$ and 12 GeV/c (here $p_{lab}=p_{1}$ is the incident
beam momentum)
as function of $\vec{p}_{m}$ for the  two transverse and
the longitudinal directions of the missing momentum, respectively.
Besides the results
for the whole IFSI factor (\ref{eq:1.19}) (solid curve) corresponding
to
the inclusive reaction, we also show the
results for the exclusive reaction
obtained with the coherent IFSI factor (\ref{eq:1.22})
(dashed curve).
To illustrate the role of the interference
and shadowing effects in the incoherent IFSI we present in Fig.~1-3 the
results obtained
keeping in the square brackets in the right-hand side of
Eq.~(\ref{eq:1.19}) only the terms $G_{i}$, $G^{i}$ and
the diagonal second order terms $G_{i}^{i}$ (dot-dashed curve).
In Figs.~1-3 we also show the nuclear transparency for the
exclusive reaction obtained neglecting
the shadowing term $G_{34}$
in (\ref{eq:1.22p}).
Figs.~1-3 demonstrate that the IFSI-distortion effects
are strong both for the inclusive and exclusive reactions.
It is seen that the increase of $p_{lab}$ from 6 up to
12 GeV/c practically does not change the nuclear transparency.
However, as one can see the $\vec{p}_{m}$-dependence
of the nuclear transparency appears to be
sensitive to the shell structure of the target nucleus.
Figs.~1,~2 show that for the both transverse
directions of the missing momentum
the incoherent IFSI become very important at
$p_{m\perp}\gsim 150-200$ MeV/c. In the case of the parallel
kinematics the difference between the nuclear transparencies for
the inclusive and exclusive reactions is relatively small
at $|p_{m,z}|\lsim 200$ MeV/c.
The results presented in
Figs.~1-3 indicate that in the case of $(p,2p)$ reaction
in the region of $p_{m}\lsim 100-150$ MeV/c the
incoherent IFSI increase the nuclear transparency by 5-10\%.
The Glauber analysis of $(e,e'p)$ reaction \cite{NSZ}
yields the relative effect of the incoherent rescatterings
$\lsim 3$\% in this region of the missing momentum. Thus, we see
that the transition from $(e,e'p)$ to $(p,2p)$ reaction
enhances
the effect of the incoherent rescatterings
substantially. None the less our
analysis indicates that the optical approach which neglects the incoherent
rescattering
is a reasonable starting point for qualitative evaluation of the nuclear
transparency in $(p,2p)$ reaction for the relatively small
missing momenta ($p_{m}\lsim 100-150$ MeV/c).
This fact is of great importance for the prospect of the further study
of CT effects in this reaction. From this point of view it
is also important that, as one can see from Figs.~1-3,
the shadowing correction also does not
affect the nuclear transparency for exclusive reaction
considerably, and can be neglected
in the above region of the missing momentum.
Notice, that
Figs.~1-3 demonstrate that
neglecting the shadowing and interference effects in the incoherent
IFSI
leads to a considerable underestimate of the contribution
of the incoherent
rescatterings.

In Figs.~4-6 we present the nuclear transparency calculated for the
kinematical domains including all the values of the two components
of the missing momentum. The legend of the curves is the same as in
Figs.~1-3. It is seen that the integration
over the two components of the missing momentum
enhances the relative effect of the incoherent IFSI as compared with
the unintegrated nuclear transparency. The contribution of the
incoherent rescatterings becomes important even in the case of
the $p_{m,z}$-dependence of $T_{A}$.

The results for the integrated nuclear transparency obtained for
the same versions of the IFSI factors as in Figs.~1-6 are presented
in Fig.~7. Fig.~7 further demonstrates that neglecting the shadowing
and interference terms leads to the underestimate of
the contribution of the
incoherent rescatterings in the inclusive $(p,2p)$ scattering
by the factor $\sim 2$. In the
case of the exclusive reaction the shadowing in the final
$pp$-system gives rise to the increase of  $T_{A}$ by $\sim 10$\%.

The curves shown in Fig.~6 were obtained according to the formula
(\ref{eq:1.25p}). In ref. \cite{BNL1} the
$p_{m,z}$-dependence of the nuclear transparency were extracted
from the cross section of $(p,2p)$ scattering
making use of for the one-dimensional SPMD the experimentally
measured $p_{m.y}$-distribution (after its normalization
to unity). For this reason the data of
\cite{BNL1} must be compared with the theoretical nuclear transparency
defined as
\beq
T_{A}^{BNL}(p_{m,z})=\frac{w_{z}(p_{m,z})}{w_{y}^{norm}(p_{m,z})}\,,
\label{eq:2.2}
\eeq
where
\beq
w_{y}^{norm}(p)=\frac{w_{y}(p)}{T_{A}}
\label{eq:2.3}
\eeq
is the normalized to unity one-dimensional IFSI-distorted
$p_{m,y}$-distribution.
The comparison of the theoretically calculated ratio
(\ref{eq:2.3}) for the whole IFSI factor (\ref{eq:1.19})
with the experimental data of ref. \cite{BNL1} for
$^{27}Al(p,2p)$ scattering  is presented
in Fig.~8.
As one can see, the predictions of the Glauber model
are in good agreement with the data \cite{BNL1} at $p_{lab}=6$ GeV/c.
Notice, that the fall down of the theoretically calculated
nuclear transparency with increase of $|p_{m,z}|$ is a consequence of
the above mentioned enhancement of the contribution of the
incoherent rescatterings in the region of large $|p_{m,y}|$ as
compared with the case of large $|p_{m,z}|$.
At $p_{lab}=10$ the experimental values of the transparency
in the region of  $p_{m,z}>0$ are in excess of the
theoretical curve by the factor $\sim 2$. At $p_{lab}=12$ GeV/c
only one experimental
point  (the bin
$200<p_{m,z}<300$ MeV/c) overshoots the Glauber curve.

In Fig.~9 we compare the theoretical
$p_{m,y}$-distribution (\ref{eq:2.3})
with the BNL experimental data \cite{BNL2} for $^{12}C(p,2p)$ and
$^{27}Al(p,2p)$ reaction at $p_{lab}=6$ and 10 GeV/c.
Besides the predictions obtained with the whole IFSI factor
(solid curve) we show the distribution for the exclusive
reaction and the one-dimensional SPMD. As one can see
the exclusive distribution and SPMD differ drastically from
the experimental distribution. The theoretical distribution
obtained with the whole IFSI factor is in good agreement
with experimental data for both nuclei at $p_{lab}=6$
GeV/c. However, at $p_{lab}=10$ GeV/c for $^{27}Al(p,2p)$
reaction the width of the experimental distribution
is in excess of the width of the theoretical distribution.
In all probability the disagreements of the Glauber model
predictions in the cases of the $p_{m,z}$-dependence of the nuclear
transparency, Fig.8, and the $p_{m,y}$-distribution, Fig.9, with the
data of ref. \cite{BNL1,BNL2} for $^{27}Al(p,2p)$ reaction
at $p_{lab}=10$ GeV/c are caused by the same reason.

\section{Conclusions}
The purpose of this work has been a study of
IFSI effects in
hard $(p,2p)$ scattering in the region
of moderate missing momenta $|\vec{p}_{m}|\lsim 300$ MeV/c
within the Glauber model.
To perform such an analysis, we generalized the Glauber theory
developed for the small-angle hadron-nucleus collisions at high energy,
to the case of hard $(p,2p)$ reaction.
We studied the missing momentum dependence
of IFSI effects both for inclusive and exclusive $(p,2p)$ scattering.
The analysis was performed
taking into account the shadowing and interference effects
in IFSI which were not discussed previously.

Our numerical results show that the missing momentum distribution
in $(p,2p)$ reaction is substantially affected by IFSI
as compared to the PWIA case both for the inclusive
and exclusive conditions.
In the inclusive
$(p,2p)$ scattering the incoherent IFSI become dominant at
$p_{m\perp}\gsim 150-200$ MeV/c.
Our results show that
neglecting the shadowing
and the interference effects leads to considerable
underestimation of the contribution of the incoherent rescatterings.
In the region of
$p_{m}\lsim 100-150$ MeV/c the incoherent IFSI increase the
missing momentum distribution only by 5-10\%.
Therefore, at the relatively small missing
momenta the optical approach is still applicable for
a qualitative evaluation of IFSI effects. This fact is of
great importance for the further investigations of CT
effects in $(p,2p)$ reactions
because neglecting the incoherent rescatterings simplifies
the calculations within the coupled-channel approach considerably.
Our results also indicate
that in future experiments it is highly desirable to
measure the nuclear transparency separately for the
regions of small ($p_{m}\lsim 150$ MeV/c) and high
($p_{m}\gsim 200$ MeV/c) missing momenta.

For the first time nuclear transparency measured
in the BNL experiment \cite{BNL1} has been compared with the theoretically
calculated transparency
defined according to the prescription of ref. \cite{BNL1}.
We emphasize that for the strong distortions of $W_{y}(p_{m,y}),
W_{z}(p_{m,z})$ and the strong dependence of $T_{A}^{BNL}(p_{m,z})$
on $p_{m,z}$ at fixed beam momentum $p_{lab}$, it does nor make any
sense to plot $T_{A}^{BNL}$ vs. $p_{lab}^{eff}=p_{lab}(1-{p_{m,z}
\over m_{p}})$ and it is erroneous
to interpret the so obtained $p_{lab}^{eff}$
dependence of $T_{A}^{BNL}$
as the true energy dependence of nuclear transparency.
We also compared the predictions of the Glauber model with
the $p_{m,y}$-distribution observed in the BNL experiment \cite{BNL2},
such a comparison has also been performed for the first time.
In both cases a good agreement of the
Glauber model predictions with experiment was
found at $p_{lab}=6$ GeV/c.
However, the predictions of the Glauber model disagree
with the data of refs. \cite{BNL1,BNL2} at $p_{lab}=10$ GeV/c.
Our analysis indicates that the discrepancy between the
Glauber model results for
nuclear transparency and the $p_{m,y}$-distribution
and the data from the BNL experiment
at $p_{lab}=10$ GeV/c are likely to be connected with the same cause.

Finally it is appropriate to comment on the status of predictions
of the one-channel
Glauber model and on the role of the off-diagonal rescatterings
in the GeV's energy region.
 There is a widespread opinion that
contribution of the off-diagonal rescatterings
vanishes
at the energies much smaller then
the CT energy scale
$E_{CT}\sim R_{A}(m_{p}^{*2}-m_{p}^{2})/2\sim 10-20$ GeV
(here $m_{p}^{*}$ is the mass of the radial excitation of the proton).
Our recent analysis \cite{EEPCT} demonstrates that
in the case of $(e,e'p)$ and $(p,2p)$ reactions this is only
the case for the coherent rescatterings, while the contribution
of the incoherent off-diagonal rescatterings becomes small
only at the energy of the fast proton(s)
$E_{p}\lsim E_{inc}\sim\Gamma_{p^{*}}m_{p^{*}}l_{int}\sim 2-3$ GeV (here
$l_{int}\sim (\sigma_{tot}(pN)\langle n_{A})^{-1}$ is
the mean free path of the proton in the nuclear medium,
$\Gamma_{p^{*}}$ is the width of the resonant state).
At higher energies ($\gsim E_{CT}$) the off-diagonal
rescatterings increase the nuclear transparency
for the coherent rescatterings and decrease it for the incoherent
rescatterings. Such an
interplay of the coherent and incoherent
IFSI can lead to an irregular energy dependence of the nuclear transparency
in inclusive $(p,2p)$ reaction in the energy region of the BNL
experiment.
The analysis of ref. \cite{EEPCT}
shows that for $(e,e'p)$ reaction the off-diagonal rescatterings
may enhance the
contribution of the incoherent rescatterings by the factor
$\sim 2$ in the energy region of the final
proton $E_{inc}\lsim E_{p}\lsim E_{CT}$. In the case of $(p,2p)$ reaction
the enhancement factor	must be $\sim 4$.
Consequently, the off-diagonal incoherent rescatterings may
increase the nuclear transparency
in $(p,2p)$ reaction at $p_{lab}\sim 10$ GeV/c by the
factor $\sim 2$. Notice, that correlation of the increase of
nuclear transparency with the broadening of $p_{m,y}$-distribution
observed in \cite{BNL1,BNL2} in $^{27}Al(p,2p)$ reaction at
$p_{lab}=10$ GeV/c
gives evidence in favor of the off-diagonal incoherent
rescatterings as a cause of the above rise of nuclear transparency.
The above estimate corresponds to the off-diagonal rescatterings
with the color-singlet intermediate $3q$ states,
and here it is appropriate to comment on the possible contribution
from the hidden-color intermediate states. Namely,
at high energies the scattering angle in the laboratory frame
$\theta_{l.f.}\ll 1$,
and the two $3q$ states produced in hard $pp$ collision can
interact with the same spectator nucleon.
For this reason the new kind of the hidden-color
(for instance the $(3q)_{\{8\}}(3q)_{\{8\}}$ state)  intermediate
states of the two final $3q$ states produced in hard
$pp$ collision may come into play.
In the case of the production of
the $(3q)_{\{8\}}(3q)_{\{8\}}$ state the intermediate hidden-color
state can undergo the transition to the normal $(3q)(3q)$ state
after rescattering on the spectator nucleon
$(3q)_{\{8\}}(3q)_{\{8\}}+N\rightarrow (3q)_{\{1\}}(3q)_{\{1\}}+N$.
It is important that in contrary to
the discussed in the present paper shadowing effect in
the final $pp$ system, the transition of the hidden-color
$(3q)_{\{8\}}(3q)_{\{8\}}$ state into the normal
$(3q)(3q)$ state requires only
one Pomeron exchange.
Of course, the production amplitude
for the hidden-color states will be suppressed by the Sudakov
form factor. However, in the GeV's energy region this
mechanism may be potentially important due to the enhancement by
the factor $\sim N_{c}^{2}$ as compared with the production
of the normal $(3q)(3q)$ states in hard $pp$ interaction.
Thus, we see that
one can expect a complicated interplay of the diagonal and off-diagonal
rescatterings
in $(p,2p)$ reaction in GeV's energy region.
For this reason, we regard
the predictions of the Glauber model only as a baseline which
allows us to understand the gross features of the coherent
and incoherent IFSI and can help in further investigation
of CT effects in $(p,2p)$ reaction.
\vspace{.2cm}\\
\acknowledgements

 This work was partly supported by the
 Grant N9S000 from the International Science Foundation and
 the INTAS grant 93-239.
AAU acknowledges Prof. A. Zichichi and ICSC- World Laboratory
for financial
support.
BGZ wishes to gratefully acknowledge the hospitality of the
Interdisciplinary Laboratory of SISSA and the Institut
f\"ur Kernphysik, KFA, J\"ulich.

\pagebreak

\pagebreak
{\large \bf Figure captions:}
\begin{itemize}

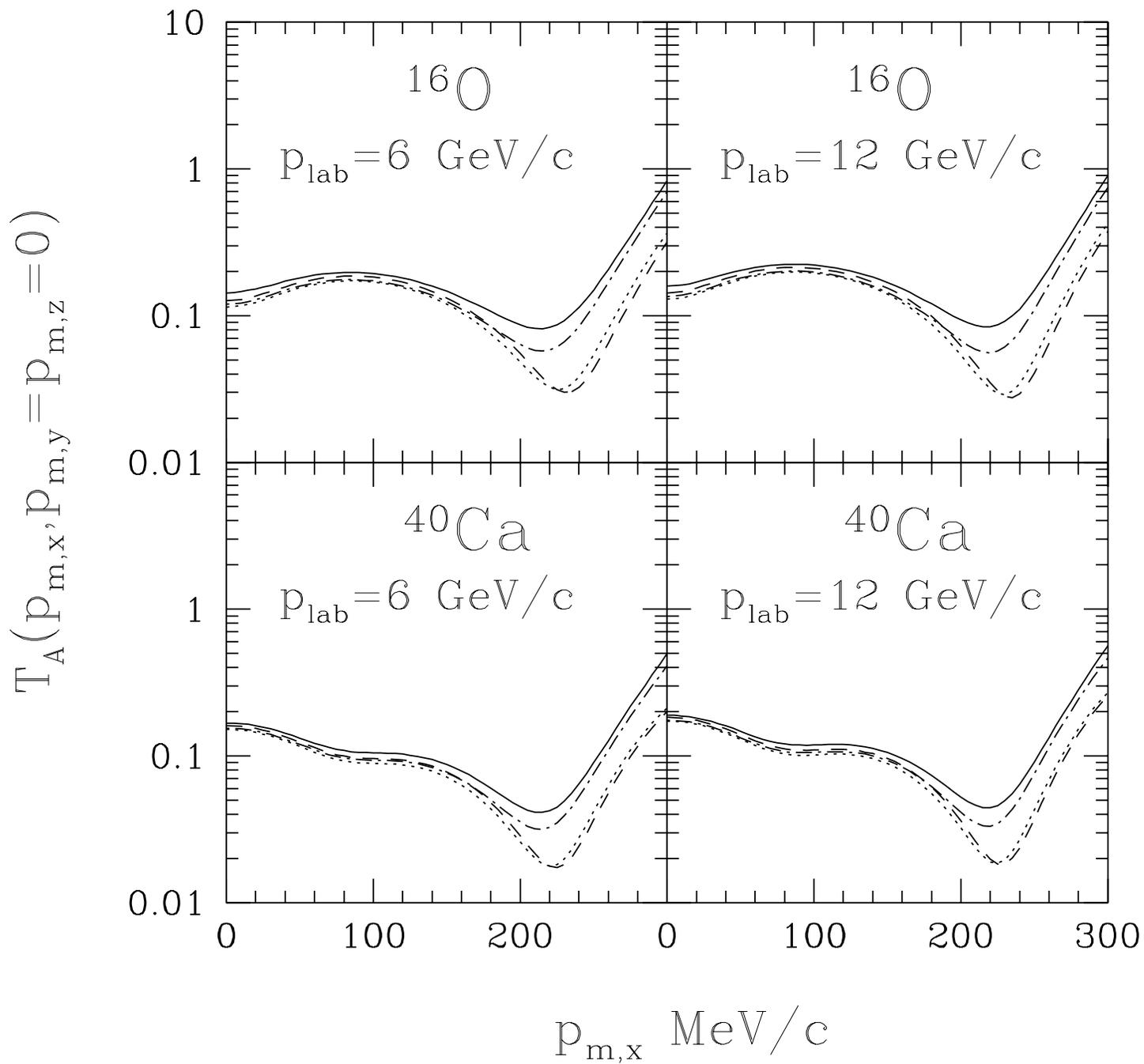
\item[Fig.~1]

~- The $p_{m,x}$-dependence of nuclear transparency for
$^{16}O(p,2p)$ and $^{40}Ca(p,2p)$ scattering
at $p_{m,y}=p_{m,z}=0$.
The solid curve is for the inclusive reaction,
the dashed curve is for the exclusive reaction,
the dot-dashed curve shows the results
obtained keeping in the IFSI factor (\ref{eq:1.19})
only the terms $G_{i}$, $G^{i}$ and $G_{i}^{i}$,
the dotted curve corresponds to the IFSI factor
(\ref{eq:1.22}) without the shadowing term $G_{34}$
in (\ref{eq:1.22p}).

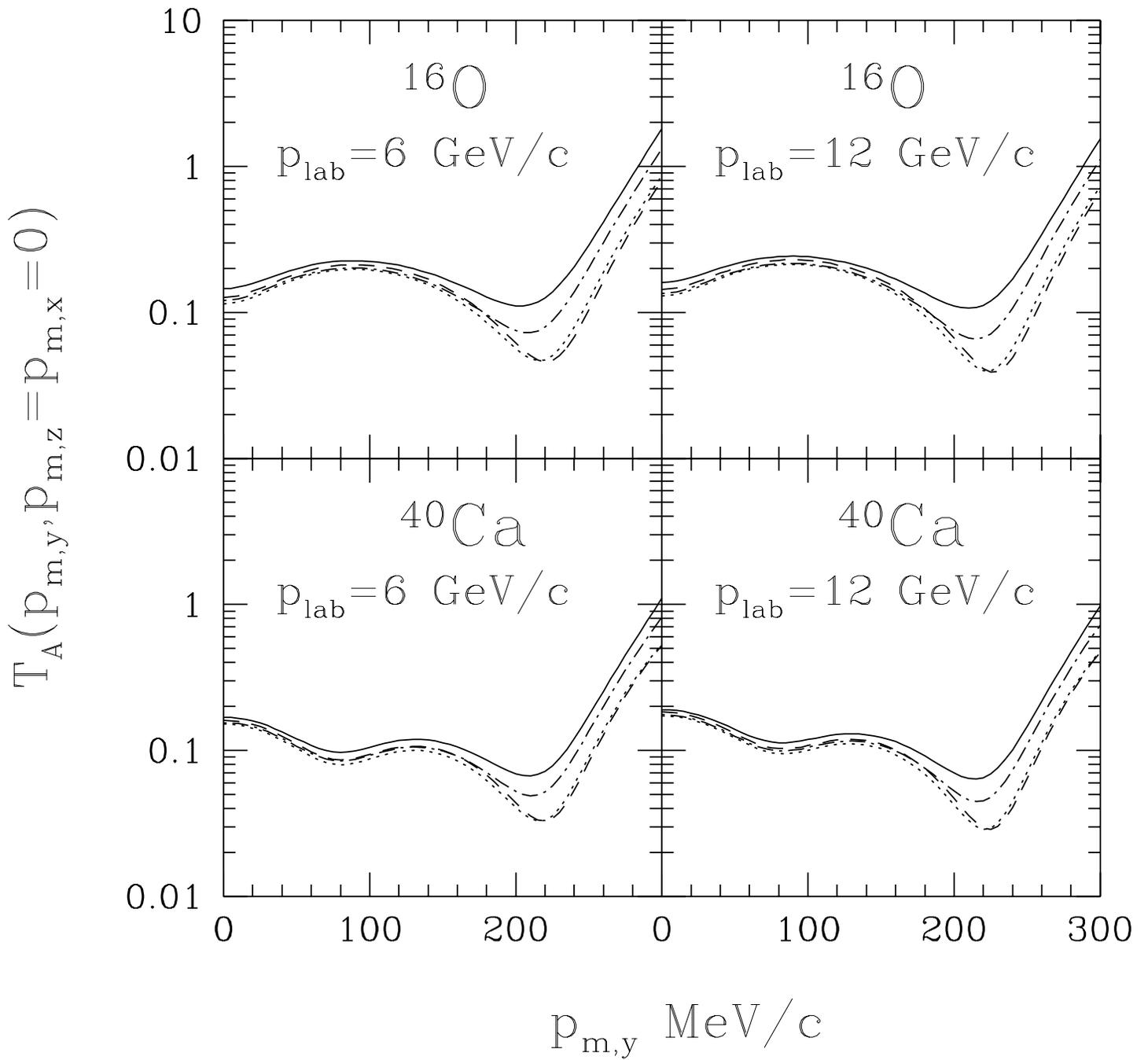
\item[Fig.~2]

~- The same as Fig.~1, but for the $p_{m,y}$-dependence
of nuclear transparency at $p_{m,z}=p_{m,x}=0$.

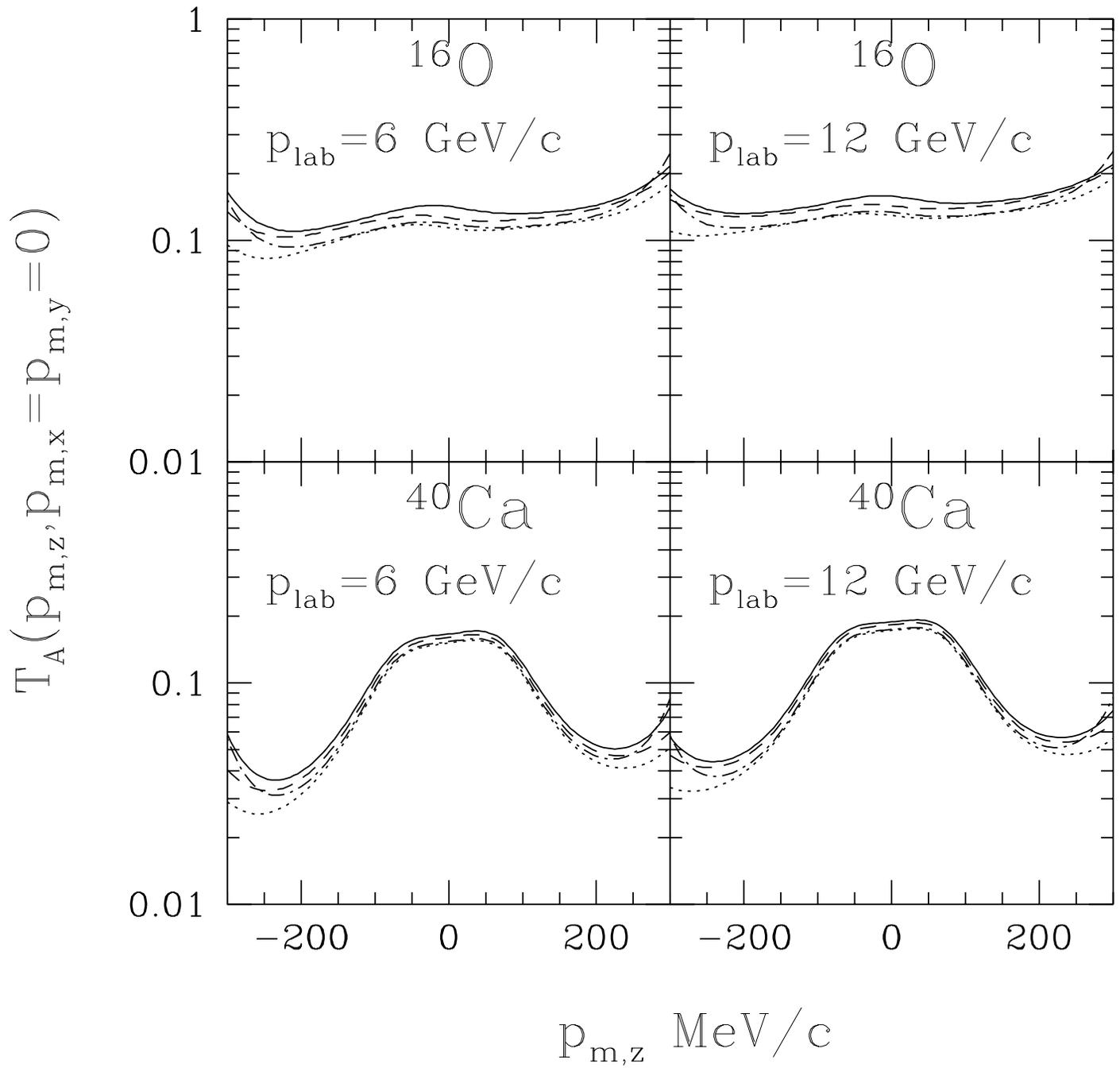
\item[Fig.~3]

~- The same as Fig.~1, but for the $p_{m,z}$-dependence
of nuclear transparency at $p_{m,x}=p_{m,y}=0$.

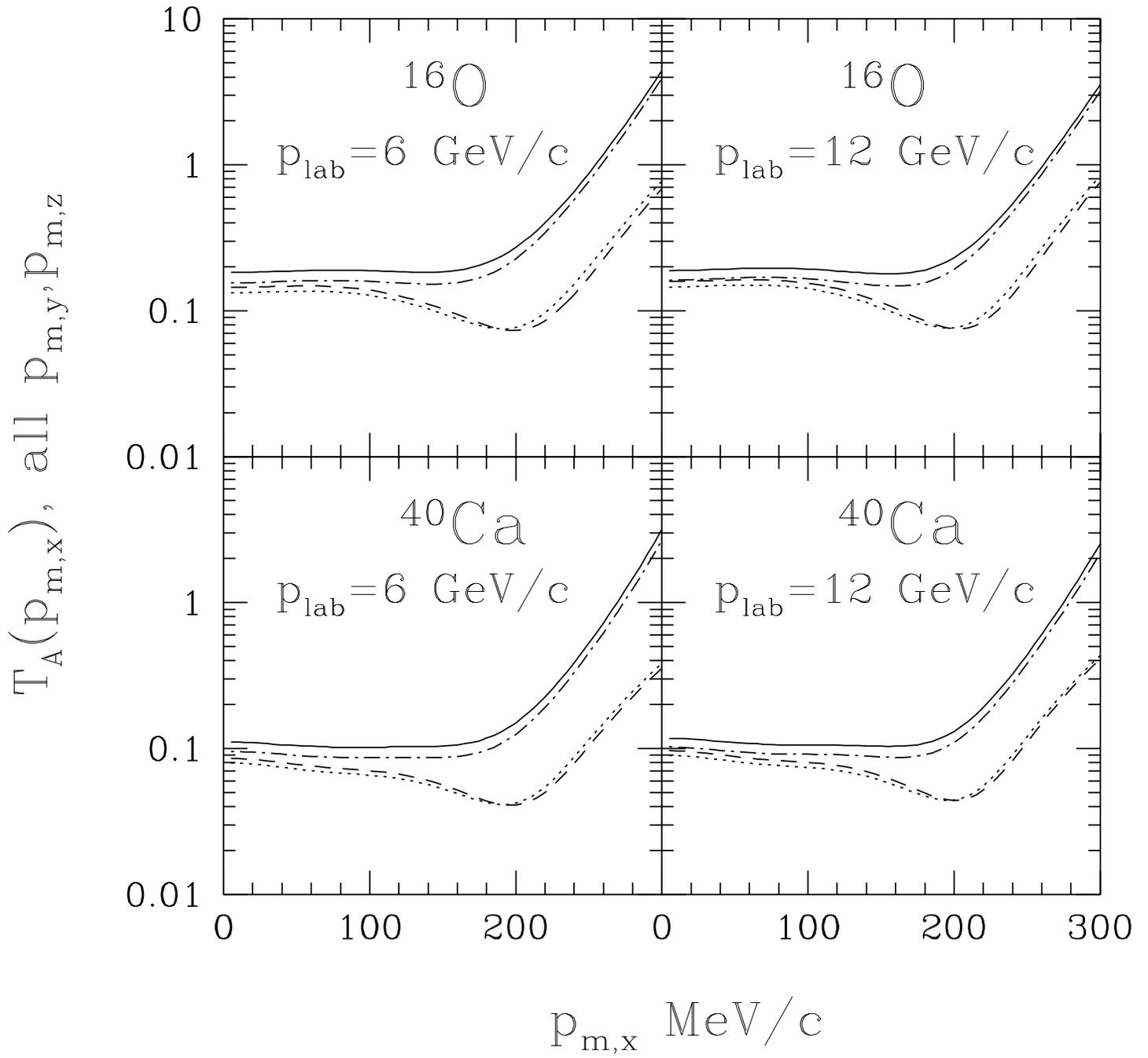
\item[Fig.~4]

~- The $p_{m,x}$-dependence of nuclear
transparency integrated over $p_{m,y}$ and $p_{m,z}$.
The legend of the curves is the same as in Figs.~1-3.

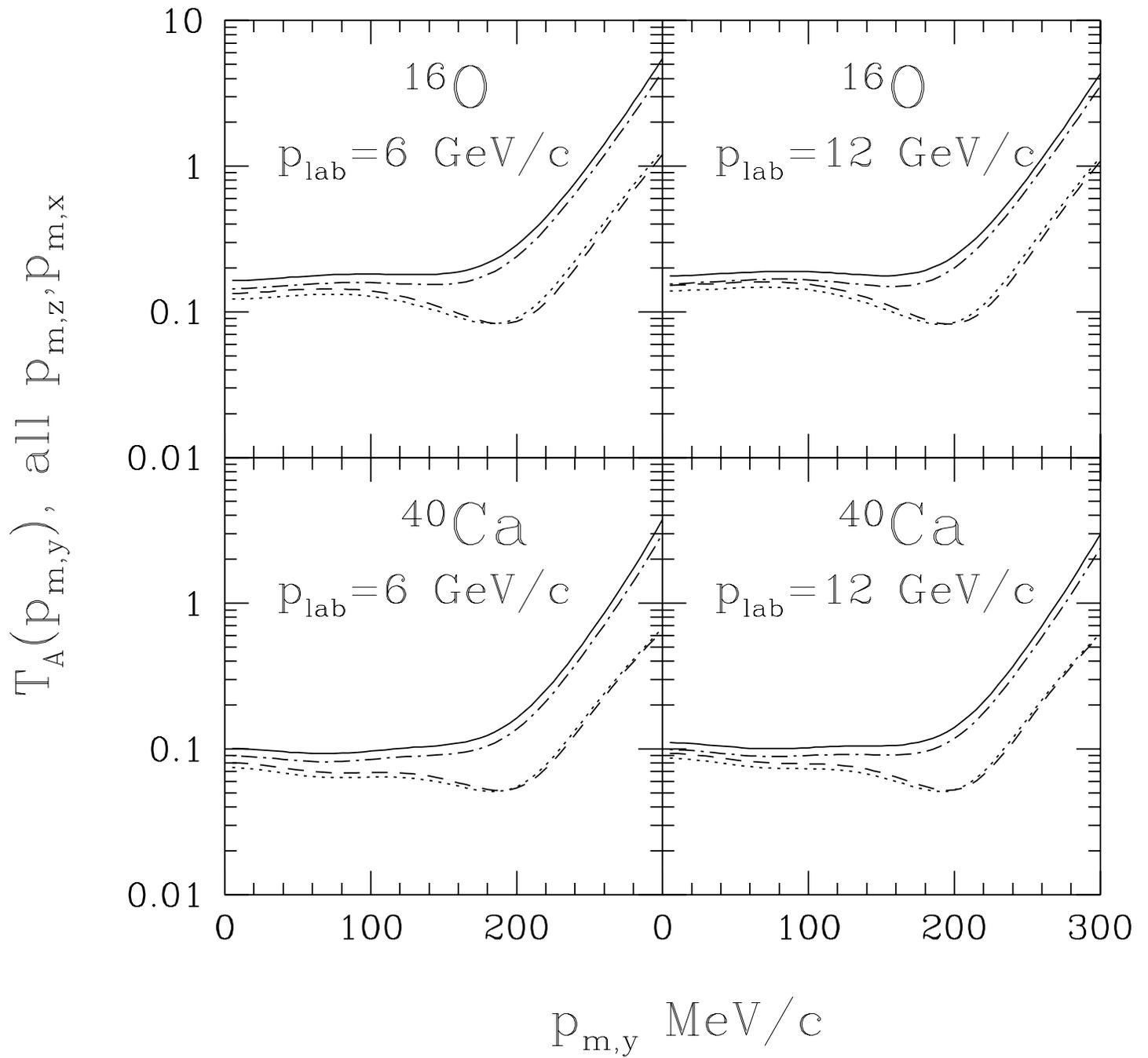
\item[Fig.~5]

~- The $p_{m,y}$-dependence of nuclear
transparency integrated over $p_{m,z}$ and $p_{m,x}$.
The legend of the curves is the same as in Figs.~1-3.

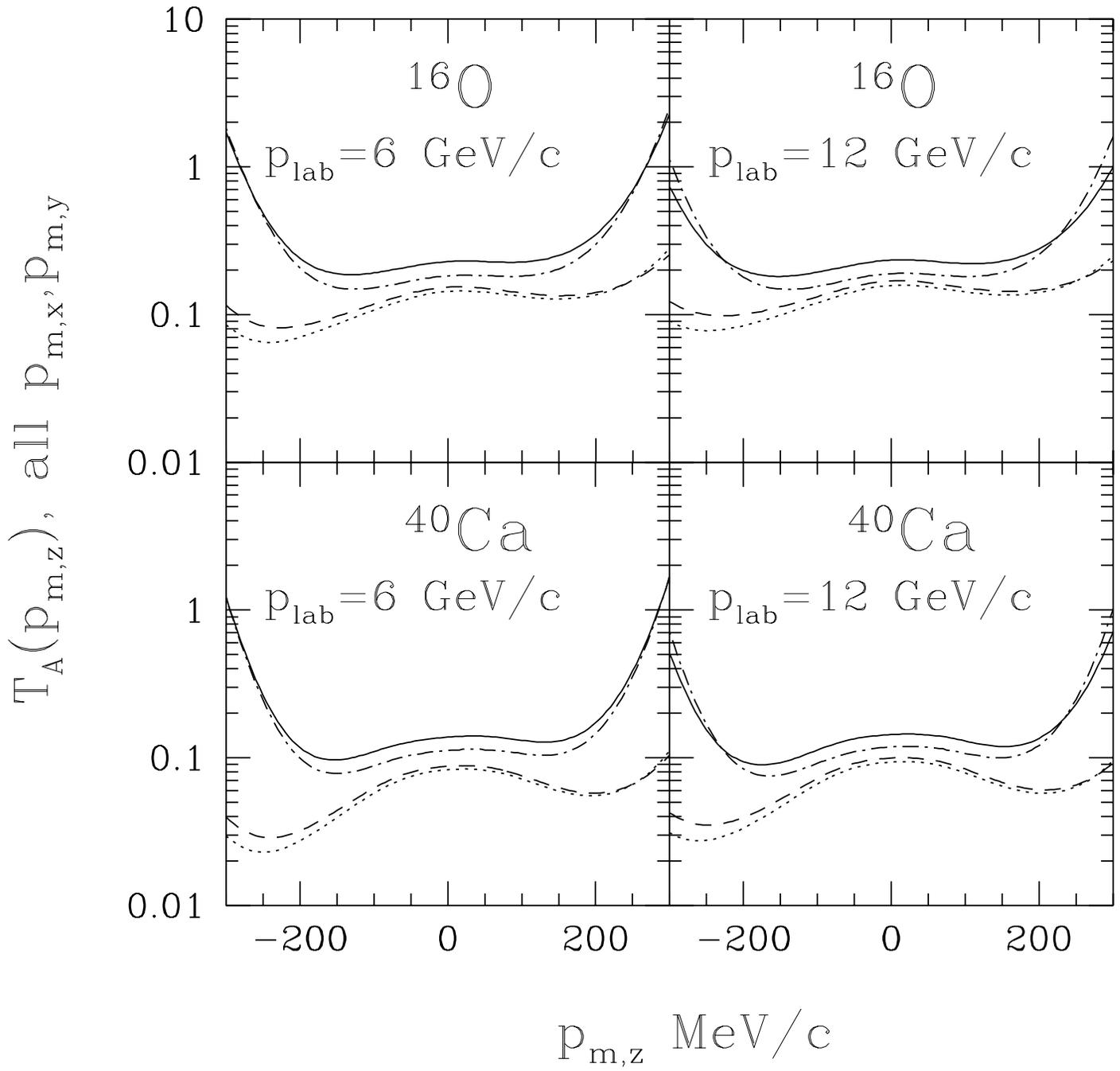
\item[Fig.~6]

~- The $p_{m,z}$-dependence of nuclear
transparency integrated over $p_{m,x}$ and $p_{m,y}$.
The legend of the curves is the same as in Figs.~1-3.

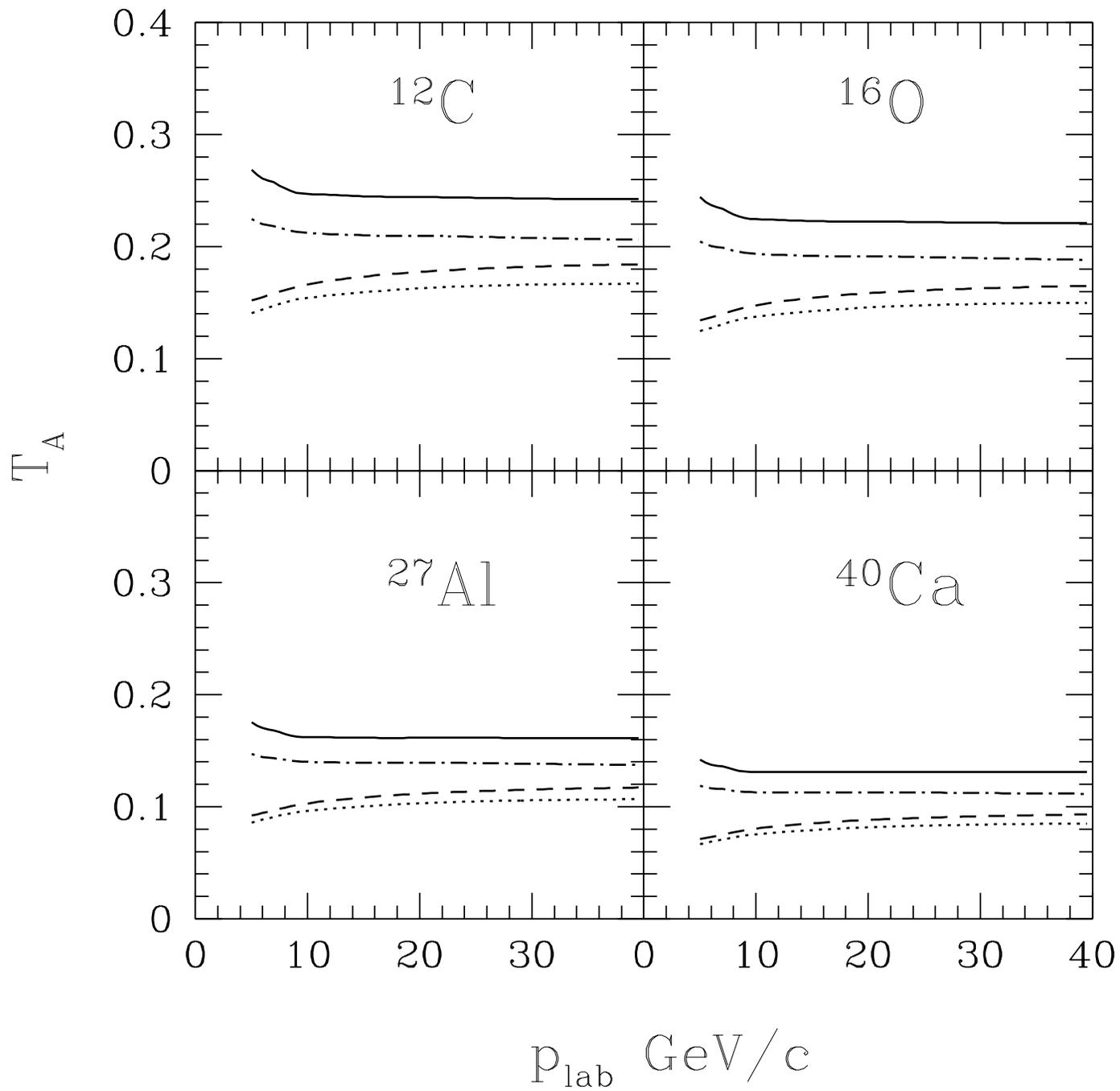
\item[Fig.~7]

~- Integrated nuclear transparency for
$^{12}C(p,2p)$, $^{16}O(p,2p)$, $^{27}Al(p,2p)$ and
$^{40}(p,2p)$ reactions.
The legend of the curves is the same as in Figs.~1-3.

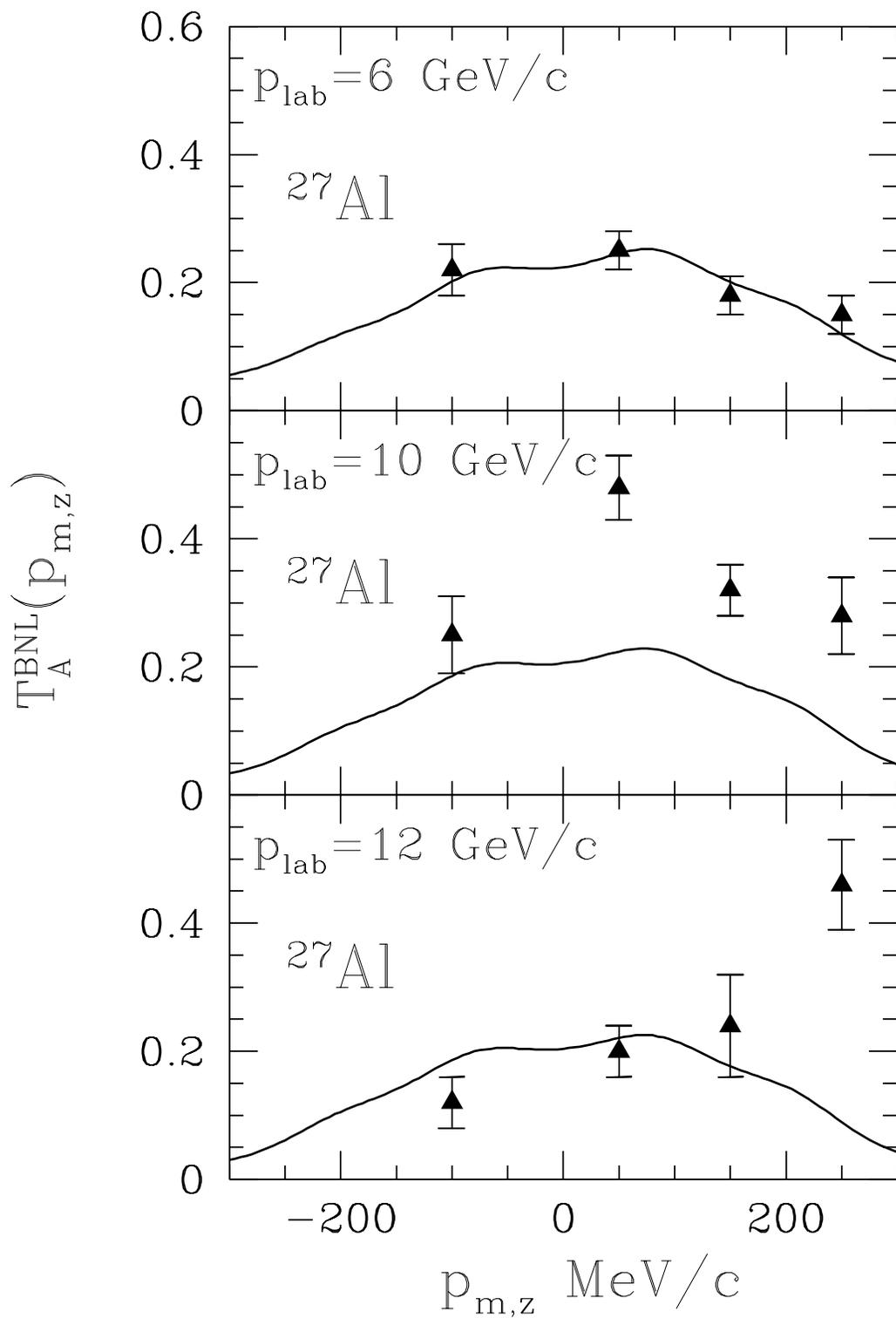
\item[Fig.~8]

~- Comparison of the Glauber model predictions
for $p_{m,z}$-dependence of nuclear transparency
defined according to Eq.~(\ref{eq:2.2}) with the
experimental data of ref. \cite{BNL1} for
$^{27}Al(p,2p)$ reaction.

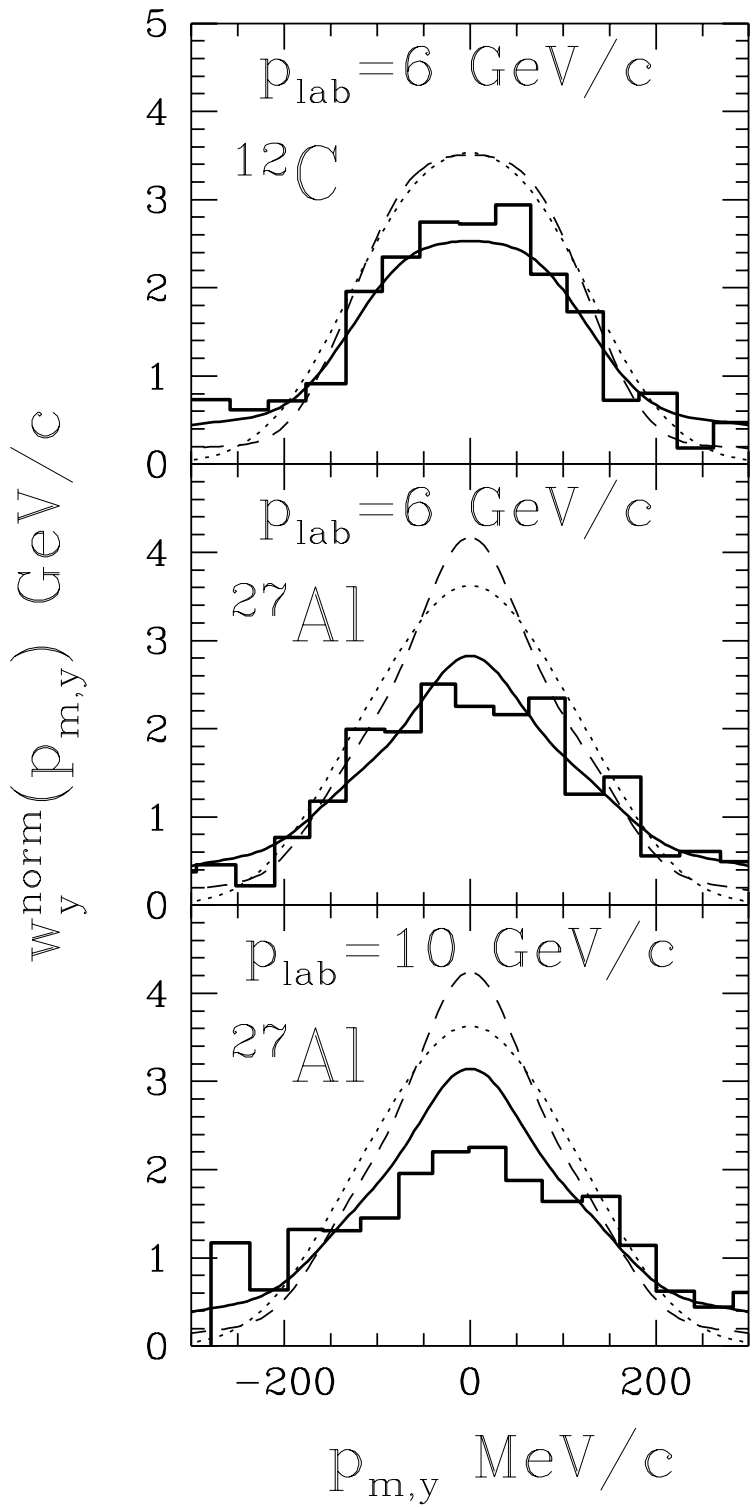
\item[Fig.~9]

~- The $p_{m,y}$-dependence of the one-dimensional normalized to unity
missing momentum distribution in $^{12}C(p,2p)$ and $^{27}Al(p,2p)$
reactions. The solid curve is for the Glauber model predictions
for the whole IFSI factor (\ref{eq:1.19}), the dashed curve shows
the predictions of the Glauber model obtained
with the coherent IFSI factor (\ref{eq:1.22}), the dotted curve shows SPMD,
the histogram shows the experimental distribution
observed in \cite{BNL2}.
\end{itemize}
\end{document}